\documentclass[12pt]{article}

\begin{document}

\begin{titlepage}

\begin{flushright}
IUHET-476\\
\end{flushright}
\vskip 2.5cm

\begin{center}
{\Large \bf Infinitesimally Nonlocal Lorentz Violation}
\end{center}

\vspace{1ex}

\begin{center}
{\large B. Altschul\footnote{{\tt baltschu@indiana.edu}}}

\vspace{5mm}
{\sl Department of Physics} \\
{\sl Indiana University} \\
{\sl Bloomington, IN 47405 USA} \\

\end{center}

\vspace{2.5ex}

\medskip

\centerline {\bf Abstract}

\bigskip

We introduce a new Lorentz-violating modification to a scalar
quantum field theory. This
interaction, while super-renormalizable by power counting, is fundamentally
different from the interactions previously considered within the
Lorentz-violating standard model extension. The Lagrange density is nonlocal,
because of the presence of a Hilbert
transform term; however, this nonlocality is also very weak. The theory has
reasonable stability and causality properties and, although the
Lorentz-violating interaction possesses a single vector index, the theory is
nonetheless CPT even. As an application, we analyze
the possible effects of this new form of Lorentz violation on neutral meson
oscillations. We find that under certain circumstances, the interaction may lead
to quite peculiar sidereal modulations in the oscillation frequency.

\bigskip

\end{titlepage}

\newpage

Recent work has stimulated a great deal of interest in the possibility of there
existing small Lorentz- and CPT-violating corrections to the standard
model. Such violations of fundamental symmetries may arise as part of the
low-energy behavior of the novel physics of the Planck scale.
The general local Lorentz-violating standard model extension (SME) has been
developed~\cite{ref-kost1,ref-kost2,ref-kost12}, and the
stability~\cite{ref-kost3} and
renormalizability~\cite{ref-kost4} of this extension have been studied. The
minimal SME includes superficially renormalizable operators that are
invariant under the standard model gauge group.

The SME provides a good
framework within which to analyze the
results of experiments testing Lorentz violation. To date, such
experimental tests have included studies of matter-antimatter asymmetries for
trapped charged particles~\cite{ref-bluhm1,ref-bluhm2,ref-gabirelse,
ref-dehmelt1} and bound state systems~\cite{ref-bluhm3,ref-phillips},
determinations of muon properties~\cite{ref-kost8,ref-hughes}, analyses of
the behavior of spin-polarized matter~\cite{ref-kost9,ref-heckel},
frequency standard comparisons~\cite{ref-berglund,ref-kost6,ref-bear},
measurements of neutral meson
oscillations~\cite{ref-kost10,ref-kost7,ref-hsiung,ref-abe},
polarization measurements on the light from distant galaxies~\cite{ref-carroll1,
ref-carroll2,ref-kost11}, and others.

However, there are other operators, beyond those considered in the SME,
that might prove important in the ultimate low-energy effective
field theory describing Lorentz violation.
By considering theories with Lorentz violation, we are relaxing the usual
conditions that one places on the Lagrangian of a quantum field theory. It is
therefore natural to consider, in conjunction with Lorentz violation, slight
relaxations of other standard conditions, such as locality. In fact, nonlocality
is already present in a renormalizable field theory that is regulated with a
momentum cutoff, because the action of such a regulator is itself nonlocal.
Nonlocality has also previously been
suggested as a potentially desirable property of high-energy Lorentz-violating
theories~\cite{ref-kost3}.

We shall consider a particular
nonlocal operator that may be added to a Lorentz invariant Lagrangian. The
nonlocality in this operator possesses an inherently Lorentz-violating
structure. However, that nonlocality is also very weak---``infinitesimal," in
a particular sense that we shall describe.
It is also significant that the theory we
shall consider, although it possesses a nonlocal Lagrangian, does not appear
to violate any causality conditions.

Our infinitesimally nonlocal variety of Lorentz violation involves the
appearance of the Hilbert transform, which, in one dimension, takes the form
\begin{equation}
\label{eq-Hilbert}
Hu(x)={\cal P}\frac{1}{\pi}\int d\xi\frac{u(\xi)}{\xi-x},
\end{equation}
where ${\cal P}$ denotes the principal value of the integral. In the Fourier
transform
domain, the Hilbert transform generates a $\frac{\pi}{2}$ phase shift
(i.e.\ $H\sin kx=\cos kx$ and $H\cos kx=-\sin kx$ if $k>0$). When coupled with a
derivative, this phase shift can give rise to very interesting dispersion
relations, according to
\begin{equation}
\label{eq-absk}
\partial_{x}(He^{ikx})=H(\partial_{x}e^{ikx})=-|k|e^{ikx}.
\end{equation}
Dispersion relations based on (\ref{eq-absk}) appear in other areas of physics,
most notably in the Benjamin-Ono equation in fluid
mechanics~\cite{ref-benjamin},
\begin{equation}
\label{eq-benjamin}
u_{t}+c_{1}u_{x}+c_{2}uu_{x}+c_{3}Hu_{xx}=0.
\end{equation}
The Hilbert transform also arises in signal processing and in the study of
complex dispersion relations.

In the context of quantum field theory, we may introduce a similar
Lorentz-violating modification of the dispersion relation. The Lagrange
density appropriate to a scalar field theory incorporating such a modification
is
\begin{equation}
\label{eq-LH}
{\cal L}_{H}=\frac{1}{2}
(\partial^{\mu}\phi)(\partial_{\mu}\phi)-\alpha a_{H}^{\mu}\left[\phi\left(
H_{\hat{a}_{H}}\partial_{\mu}\phi\right)\right]
-\frac{m^{2}}{2}\phi^{2}.
\end{equation}
To avoid any complexities associated with the appearance of nonstandard time
derivatives, we shall take $a_{H}^{\mu}$ to be purely spacelike in the frame in
which the theory is quantized. Otherwise, we would have a Lagrangian that
would be nonlocal
in time, and this is clearly inadmissible in canonical quantum theory.
(Moreover, if a timelike component for $a_{H}^{\mu}$ were allowed, we would
need to define the theory in Wick-rotated Euclidean space.)
$H_{\hat{a}_{H}}$ represents the Hilbert transform
taken along the direction of the unit vector $\hat{a}_{H}=\vec{a}_{H}/
|\vec{a}_{H}|$,
\begin{equation}
\label{eq-Ha}
H_{\hat{a}_{H}}\phi(t,\vec{x})={\cal P}\frac{1}{\pi}\int d\xi\frac{\phi[t,
\vec{x}-(\hat{a}_{H}\cdot\vec{x})\hat{a}_{H}+\xi]}{\xi-\hat{a}_{H}\cdot\vec{x}}.
\end{equation}
$\alpha$ is either $\pm1$ and determines the sign of the novel term.
The interactions we are considering are by no means the most general that may be
represented by the Hilbert transform. However, (\ref{eq-LH}), which contains
``collinear" operators $\vec{a}_{H}\cdot\vec{\partial}$ and $H_{\hat{a}_{H}}$,
represents the most natural of the possible Lagrangians that could be
considered.

It also possible to envision a reasonable scenario in which an interaction
such as $a^{\mu}_{H}
\left[\phi\left(H_{\hat{a}_{H}}\partial_{\mu}\phi\right)\right]$ could arise in
an effective field theory. The Benjamin-Ono equation (\ref{eq-benjamin})
describes the waves and solitons that occur within a certain regime of fluid
mechanics. The underlying equations of motion (i.e.\ the Navier-Stokes
equations) do not involve Hilbert transforms, but the effective theory does. We
envision something analogous in the particle physics situation. For example, if
the elementary particles we observe are actually the solitary waves of some
underlying theory, then, as in the Benjamin-Ono case, the dynamics of these
solitary waves could be determined by interactions involving Hilbert
transforms. In the fundamental theory, the absolute value appearing in the
dispersion relation could be replaced by a smooth function, differing from
the absolute value only in a region of momentum space whose size is suppressed
by some power of a large mass scale. The coefficient $a^{\mu}_{H}$ could arise
in the same way as any other Lorentz-violating parameter, perhaps as the vacuum
expectation value of a vector-valued field.

The theory with $a^{\mu}_{H}$
is closely related to a simpler Lorentz-violating theory, defined by
\begin{equation}
\label{eq-La}
{\cal L}_{a}=\left(\partial^{\mu}\Phi^{*}\right)\left(\partial_{\mu}\Phi\right)+
ia^{\mu}\left[\Phi^{*}\left(\partial_{\mu}\Phi\right)-\left(\partial_{\mu}
\Phi^{*}\right)\Phi\right]-m^{2}\Phi^{*}\Phi,
\end{equation}
where $\Phi$ is a complex scalar field. For instance, the $a^{\mu}$ and
$a_{H}^{\mu}$
theories have very similar stability conditions, as we shall see.
$a^{\mu}$ and $a^{\mu}_{H}$ both have dimension (mass)$^{1}$, and so both are
superficially super-renormalizable.
However, the $a^{\mu}$ theory is significantly simpler in a number of ways.

Equation (\ref{eq-absk}) tells us that the nonlocality due to the presence
of the Hilbert transform is very weak. For a wave packet that is well
localized in momentum space, the nonlocality of the dispersion relation will
not be evident, since only the Fourier modes with a single sign of
$k_{\hat{a}_{H}}\equiv\vec{k}\cdot\hat{a}_{H}$ will contribute substantially.
Seen
another way, the nonlocality only manifests itself in the neighborhood of
$k_{\hat{a}_{H}}=0$. This breakdown of nonlocality is analogous to the breakdown
of associativity that occurs in the presence of a magnetic monopole, when the
Jacobi identity fails at a single point in
space~\cite{ref-jackiw,ref-grossman,ref-wu}. Since the breakdown of the theory's
expected properties occurs only
on a set of measure zero in each case, the theory can
remain physically reasonable.

The discrete symmetries of the perturbation $\left[\phi\left(
H_{\hat{a}_{H}}\partial_{\mu}\phi\right)\right]$ may be easily worked out. Since
we are only considering a purely spacelike $a^{\mu}_{H}$, the derivative
operator is odd under
P and even under C and T. The Hilbert transform has the same symmetries. The
sign change
of $H_{\hat{a}_{H}}$ under parity comes from the denominator of (\ref{eq-Ha}).
The behavior under time reversal can be determined from the facts that
$H_{\hat{a}_{H}}$
does not interact with the time coordinate and that the Hilbert transform is
entirely real, so there are no sign changes arising from the antiunitarity
of T.
Charge conjugation is of course trivial in a real scalar field theory.
This means that the coefficient $a_{H}^{\mu}$ describes a Lorentz-violating
parameter with an odd number of Lorentz indices, which is nonetheless CPT-even.
There are no local Lorentz-violating interactions with this property.

Because it is even under CPT, $a_{H}^{\mu}$ cannot contribute radiatively to
any of the Lorentz-violating coefficients in the SME at leading order.
Conversely, none of the SME parameters will contribute to the renormalization of
$a_{H}^{\mu}$; any radiative corrections must be proportional to $a_{H}^{\mu}$
itself. In fact, if we endow the scalar field in (\ref{eq-LH}) with a
$\phi^{4}$ interaction, there will be no one-loop radiative corrections to
$a_{H}^{\mu}$ at all, just as there is no field strength renormalization
in the usual $\phi^{4}$ theory at
leading order. The only one-loop self-energy diagram is a tadpole, with no
dependence on the external momentum. This diagram contributes a mass
renormalization and nothing more. So the effective value of $a_{H}^{\mu}$ is not
sensitive to any lowest-order radiative effects.

The stability properties of this theory can be easily verified in the
quantization frame.
For $\alpha=1$,
the contribution to the energy from the Hilbert transform term is always
positive, so stability is assured.  If $\alpha$ is negative, then the theory
is stable provided that $|\vec{a}_{H}|\leq m$. This is essentially the same
stability condition that arises in the presence of the local $(a\cdot\partial)$
interaction from ${\cal L}_{a}$. In either case, the condition ensures that the
potentially negative Lorentz-violating contributions cannot dominate the energy.
We shall naturally consider only stable situations in this paper, and, in any
case, we expect that $\frac{|\vec{a}_{H}|}{m}$ should be very small in any
physically relevant scenario.

The question of causality is trickier. However, we can demonstrate
one important
result in this area. We consider the group velocity $\vec{v}_{g}$ for the
particles under consideration. For a wave packet well localized around
three-momentum
$\vec{k}_{0}$,
$\vec{v}_{g}=\left.\vec{\nabla}_{\!\vec{k}}\,E\left(\vec{k}\right)
\right|_{\vec{k}=\vec{k}_{0}}$.
Away from $k_{\hat{a}_{H}}=0$, the absolute value in the dispersion relation
\begin{equation}
E\left(\vec{k}\right)=\sqrt{\vec{k}^{2}+2\alpha|\vec{a}_{H}||k_{\hat{a}_{H}}|+
m^2}
\end{equation}
is not evident, and the group velocities are equivalent to those in the theory
defined by ${\cal L}_{a}$. The ${\cal L}_{a}$ theory possesses full
microcausality; this is verified explicitly in~\cite{ref-kost3} for a fermionic
version of the $a^{\mu}$ theory, related to the bosonic one by
supersymmetry~\cite{ref-berger}.
In the vicinity of $k_{\hat{a}_{H}}=0$, the group
velocity in the $a_{H}^{\mu}$ theory ceases to be a well-defined concept;
however, while $\partial E/
\partial k_{\hat{a}_{H}}$ is discontinuous, its magnitude is bounded by unity
(provided,
of course, that $|\vec{a}_{H}|\leq m$).

The discontinuous behavior of the group velocity in this system has some
interesting effects. For a wave packet moving in the plane perpendicular to
$\vec{a}_{H}$, the momentum spread in the $\hat{a}_{H}$-direction is
centered around $k_{\hat{a}_{H}}=0$. The components of the wave packet with
${\rm sgn}\left(k_{\hat{a}_{H}}\right)>0$ will have a small velocity in the
$\hat{a}_{H}$-direction of $\frac{\alpha}{m}
\vec{a}_{H}$. The Fourier components with negative $k_{\hat{a}_{H}}$ will
have velocity $-\frac{\alpha}{m}
\vec{a}_{H}$. So no matter how well-localized the wave packet, there
will be a discontinuity of $\Delta\vec{v}=\frac{2}{m}\vec{a}_{H}$ in the
velocity, and this
will affect wave packet spreading. Over time, the wave packet will bifurcate,
with the two halves moving apart with velocity $\Delta\vec{v}$.

Since the form of Lorentz violation we are considering is invariant under C,
P, and T, it can be difficult to find an experimental signature for the
effects of $a^{\mu}_{H}$. However, like an $a^{\mu}$ interaction, these new
physics could cause changes in the structure of neutral meson oscillations.
CPT violations in the $K^{0}$ system have been strongly
constrained~\cite{ref-schwingenheuer,ref-harati}, but CPT-even Lorentz
violations are still possible. Moreover, similar effects are also possible for
the other neutral mesons---$D^{0}$, $B^{0}_{d}$, and
$B^{0}_{s}$~\cite{ref-kost13}.

In what follows, we shall neglect any CP-violating interactions that are
present in the kaon system, and we shall also neglect decay processes, although
either of these could be included without much difficulty.
To study the effects of $a^{\mu}_{H}$ on kaon oscillations, we must generalize
the Lagrange density (\ref{eq-LH}) to one involving a complex scalar field
$K$. The simplest generalization is
\begin{equation}
\label{eq-LK0}
{\cal L}_{K,0}=
(\partial^{\mu}K^{*})(\partial_{\mu}K)-2\alpha a_{H}^{\mu}\left[K^*\left(
H_{\hat{a}_{H}}\partial_{\mu}K\right)\right]-m^{2}K^{*}K.
\end{equation}
Since the Hilbert transform is purely real, the breakdown of the
Lorentz-violating term in terms of $K_{1}=\frac{1}{\sqrt{2}}(\phi+\phi^{*})$
and $K_{2}=\frac{1}{i\sqrt{2}}(\phi-\phi^{*})$ is
\begin{equation}
\label{eq-aHsplit}
2\alpha a_{H}^{\mu}\left[K^*\left(H_{\hat{a}_{H}}\partial_{\mu}K\right)\right]=
\alpha a_{H}^{\mu}\left[K_{1}\left(H_{\hat{a}_{H}}\partial_{\mu}K_{1}
\right)+K_{2}\left(H_{\hat{a}_{H}}\partial_{\mu}K_{2}\right)
\right].
\end{equation}

However, ${\cal L}_{K,0}$ does not cause any $K^{0}$-$\bar{K}^{0}$
oscillations. The normal oscillations are generated by a mass difference
between $K_{1}$ and $K_{2}$, and we must include this effect. There is also
potentially a small difference between the two values of $a^{\mu}_{H}$ in
(\ref{eq-aHsplit}) corresponding to the two fields $K_{1}$ and $K_{2}$. Like
the mass differences, the small differences in $a^{\mu}_{H}$ could be due to
differences in the interaction of the $K_{1}$ and $K_{2}$ particles with other
virtual species.
The ultimate Lagrange density we shall consider is therefore
\begin{eqnarray}
{\cal L}_{K} & = & \frac{1}{2}
(\partial^{\mu}K_{1})(\partial_{\mu}K_{1})+\frac{1}{2}
(\partial^{\mu}K_{2})(\partial_{\mu}K_{2})
-\frac{m^{2}_{1}}{2}K^{2}_{1}-\frac{m^{2}_{2}}{2}K^{2}_{2} \nonumber\\
& & -\alpha a_{H1}^{\mu}\left[K_{1}\left(
H_{\hat{a}_{H1}}\partial_{\mu}K_{1}\right)\right]
-\alpha a_{H2}^{\mu}\left[K_{2}\left(
H_{\hat{a}_{H2}}\partial_{\mu}K_{2}\right)\right].
\end{eqnarray}
The usual oscillations are generated by
the beat frequency $\Delta m\equiv m_{1}-m_{2}$. The leading
Lorentz-violating modifications may be controlled either by $\frac{\Delta m}{m}
|\vec{a}_{H}|$ or by $|\Delta\vec{a}_{H}|\equiv|\vec{a}_{H1}-\vec{a}_{H2}|$,
depending on whether
$\frac{\Delta m}{m}$ or $\frac{|\Delta\vec{a}_{H}|}{|\vec{a}_{H}|}$
is larger. We shall not consider the possibility that $\alpha$ may differ
between the two species, since this would represent a large
relative difference between
the corresponding particles' respective Lagrangians.

To study the oscillations, we must determine the energy of a kaon in the frame
of quantization, in which $a^{\mu}_{H}$ is purely spacelike. We then boost into
the rest frame of the particle. In the quantization frame, the energy for a
scalar particle with mass $m$, momentum $\vec{k}$, and a Lorentz-violating
parameter $a^{\mu}_{H}$ is approximately
\begin{equation}
E\approx\sqrt{m^{2}+\vec{k}^{2}}+\frac{\alpha|\vec{a}_{H}||k_{\hat{a}_{H}}|}
{\sqrt{m^{2}+\vec{k}^{2}}}.
\end{equation}
This is valid to first order in
$|\vec{a}_{H}|$, and henceforth, we shall neglect
any higher-order corrections, which
should be miniscule.
The group velocity $\vec{v}_{g}$ corresponding to the energy $E$ is then
\begin{equation}
\vec{v}_{g}=\frac{\vec{k}} {\sqrt{m^{2}+\vec{k}^{2}}} +\frac{\alpha|\vec{a}_{H}
|}{\sqrt{m^{2}+\vec{k}^{2}}}\left[{\rm sgn}(k_{\hat{a}_{H}})\hat{a}_{H}-\frac
{|k_{\hat{a}_{H}}|\vec{k}}{m^{2}+\vec{k}^{2}}\right].
\end{equation}
This corresponds to a Lorentz factor of
\begin{equation}
\gamma=\frac{1}{\sqrt{1-\vec{v}\,^{2}\!\!\!\!_{g}\,\,}}
=\frac{\sqrt{m^{2}+\vec{k}^{2}}}{m}
\left(1+\frac{\alpha|\vec{a}_{H}||k_{\hat{a}_{H}}|}{m^{2}+\vec{k}^{2}}\right).
\end{equation}
So the rest energy, $E_{0}=\gamma(E-\vec{v}_{g}\cdot\vec{k})$, is
simply
\begin{equation}
E_{0}=m+\frac{\alpha|\vec{a}_{H}||k_{\hat{a}_{H}}|}{m}.
\end{equation}

The rate of oscillations is determined by the difference between the rest
energies of the $K_{1}$ and $K_{2}$ modes. If $\frac{|\Delta\vec{a}_{H}|}
{|\vec{a}_{H}|}$ is small, then the signs of $k_{\hat{a}_{H1}}$ and
$k_{\hat{a}_{H2}}$ will almost always be the same; we shall 
for now neglect any slight deviations from this
around $k_{\hat{a}_{H}}\approx0$. With this approximation, the energy
difference is
\begin{equation}
\Delta E=m_{1}-m_{2}+\alpha\,{\rm sgn}(k_{\hat{a}_{H}})\vec{k}\cdot\left(
\frac{m_{2}\vec{a}_{H1}-m_{1}\vec{a}_{H2}}{m_{1}m_{2}}\right).
\end{equation}

The most obvious signature for this form of Lorentz violation would be sidereal
variations in the oscillation rate. We shall consider
this variation in two different regimes.
The momentum $\vec{k}$ consists of two parts---the momentum $\vec{k}_{\oplus}$
due to the motion of the earth relative to the quantization frame and the
momentum $\vec{k}_{L}$
of the kaons measured
in the laboratory frame. If $\vec{k}_{\oplus}\cdot\hat{a}_{H}
\gg\left|\vec{k_{L}}\right|$, then the
motion of the earth is the dominant effect. If the quantization frame is the
rest frame of the cosmic microwave background, then the velocity scale
corresponding to
$\vec{k}_{\oplus}$ is 365 km $\cdot$ s$^{-1}$. In this regime,
${\rm sgn}(k_{\hat{a}_{H}})={\rm sgn}(k_{\oplus}\cdot\hat{a}_{H})$ is a
constant. If the laboratory momentum $\vec{k}_{L}$
has component $k_{L,z}$ along the polar ($z$-)
axis and a component of magnitude $k_{L,xy}$ in the equatorial ($xy$-) plane,
and
$a_{Hj,z}$ and $a_{Hj,xy}$ are the corresponding components of $\vec{a}_{Hj}$,
then the time variation of $\Delta E$ is given by
\begin{eqnarray}
\Delta E & = & \left[m_{1}-m_{2}+\alpha\,{\rm sgn}(k_{\oplus}\cdot\hat{a}_{H})
\left(
\vec{k}_{\oplus}\cdot\frac{m_{2}\vec{a}_{H1}-m_{1}\vec{a}_{H2}}{m_{1}m_{2}}
+k_{L,z}\frac{m_{2}a_{H1,z}-m_{1}a_{H2,z}}{m_{1}m_{2}}\right)\right]
\nonumber\\
& & +\alpha\left[{\rm sgn}(k_{\oplus}\cdot\hat{a}_{H})
k_{L,xy}\frac{m_{2}a_{H1,xy}-m_{1}a_{H2,xy}}{m_{1}m_{2}}\right]\cos
(\omega_{\oplus}t+\psi).
\end{eqnarray}
Here, $\omega_{\oplus}$ is the earth's sidereal rotation frequency, and $\psi$
is
a phase determined by the initial conditions. In this case, the time-dependence
of the oscillation frequency is entirely sinusoidal, and the effects are
essentially indistinguishable from those generated by a
local Lorentz-violating term, because the absolute
value in the dispersion relation has no direct effect.

If, on the other hand, the magnitude of the laboratory momentum $\vec{k}_{L}$ is
much larger than $\vec{k}_{\oplus}\cdot\hat{a}_{H}$, then the situation is more
complicated, because the sign of $k_{\hat{a}_{H}}$ may change.
Let $\theta_{a_{H}}$ and $\theta_{k_{L}}$ be the colatitudes
corresponding to the directions of $\vec{k}_{L}$ and $\vec{a}_{H}$. If we
neglect any contributions from $\vec{k}_{\oplus}$, then the sign of
$k_{\hat{a}_{H}}$ will change during the course of the earth's rotation exactly
if $\cos^{2}\theta_{a_{H}}<
\sin^{2}\theta_{k_{L}}$ (i.e., if $0\leq\theta_{a_{H}}\leq\frac{\pi}{2}$, then
there will be a change in ${\rm sgn}(k_{\hat{a}_{H}})$ if $\left|\frac{\pi}{2}-
\theta_{k_{L}}\right|<\theta_{a_{H}}$). The energy difference is thus
\begin{eqnarray}
\Delta E & = & m_{1}-m_{2}+\alpha\left|\frac{k_{L,z}a_{H1,z}}{m_{1}}+
\frac{k_{L,xy}a_{H1,xy}}{m_{1}}\cos(\omega_{\oplus}t+\psi_{1})\right|
\nonumber\\
\label{eq-DeltaE}
& & -\alpha\left|\frac{k_{L,z}a_{H2,z}}{m_{2}}+
\frac{k_{L,xy}a_{H2,xy}}{m_{2}}\cos(\omega_{\oplus}t+\psi_{2})\right|.
\end{eqnarray}
The two phases $\psi_{1}$ and $\psi_{2}$ should be nearly equal; differences
between them are suppressed by further factors of
$\frac{|\Delta\vec{a}_{H}|}{|\vec{a}_{H}|}\ll 1$. If $\cos^{2}\theta_{a_{H}}
\geq\sin^{2}\theta_{k_{L}}$, the time variation of $\Delta E$ is simply
sinusoidal, as it was in the case in which $\vec{k}_{\oplus}$ dominated.
However, if $\cos^{2}\theta_{a_{H}}<\sin^{2}\theta_{k_{L}}$, then $\Delta E$ has
cusps at $\cos(\omega_{\oplus}t+\psi_{j})=-\frac{k_{L,z}a_{Hj,z}}{k_{L,xy}a_{Hj,
xy}}$. These cusps represent the most telling signature indicating the presence
of an $a_{H}^{\mu}$ interaction in the neutral kaon system. In fact, if $k_{L,z}
a_{H1,z}\approx k_{L,z}a_{H2,z}$ is small enough to be neglected, then the
oscillations in $\Delta E$
will effectively have period $\frac{\pi}{\omega_{\oplus}}$, rather than
$\frac{2\pi}{\omega_{\oplus}}$, because the time-dependent part of $\Delta E$
will be proportional to $\left|\cos(\omega_{\oplus}t+\psi)\right|$. This kind
of effect could not be generated by a simple $a^{\mu}$ coefficient.

Because the Hilbert transform interaction we have considered is even under C, P,
and T, it does not contribute directly to the quantities parameterizing the CP
and CPT violations in the meson system. It is most natural, therefore, to search
for the effects of $a_{H}^{\mu}$ through a direct analysis of the
$K^{0}$-$\bar{K}^{0}$ oscillation rate. (It is for this reason that we have
neglected any decay and CP-violating processes in our analysis, as they do not
affect this kind of analysis in any fundamental way.) A beam that initially
consists entirely of $K^{0}$ will contain, after a time $t$, fractions
$\cos^{2}\left(\frac{\Delta Et}{2}\right)$ and $\sin^{2}\left(\frac{\Delta
Et}{2}\right)$ of $K^{0}$ and $\bar{K}^{0}$ particles, respectively. By
measuring the behavior of the beam as a function of $t$, one may determine
the time dependence of its particle content. If the oscillation frequency
$\frac{\Delta E}{2}$ shows time and directional dependences of the type
displayed in (\ref{eq-DeltaE}), then this will be strong evidence for this type
of nonlocal Lorentz violation.

The current best measurements of the $K_{S}$-$K_{L}$ mass difference give
results of approximately $3.48\pm 0.01 \times10^{-6}$ eV~\cite{ref-pdg},
compared with a neutral kaon mass of 498 MeV. For KTeV kaons, with total
energies of 70 GeV, the implied sensitivity for $\frac{\Delta m}{m}
|\vec{a}_{H}|$ or $|\Delta\vec{a}_{H}|$ is of the order of $10^{-10}$ eV.
If $\frac{\Delta m}{m}|\vec{a}_{H}|$ generates the dominant contribution,
then the direct sensitivity for $|\vec{a}_{H}|$ is only at the $10^{4}$ eV
level. These potential experimental constraints are thus much less stringent
than those that can be obtained for the CPT-violating $a^{\mu}$.

In this paper,
we have presented a new possible Lorentz-violating interaction, beyond that
previously considered in the SME. This interaction, which
involves a Hilbert transform, is
superficially renormalizable, and, although it is weakly nonlocal, it could
reasonably be generated by some more complicated underlying theory. The
corresponding nonlocal theories
have causality and stability properties similar to those of
local Lorentz-violating theories. However, the Hilbert transform gives rise to
a form of Lorentz violation that is CPT even, yet which possesses a vector
index; this is
a property that has not previously been observed in the study of Lorentz
violation. We have also
considered the effects such an interaction might have on neutral meson
oscillations and shown that, when the laboratory momentum is large and aligned
appropriately, this interaction could modify the usual oscillations in a unique
way. Finally,
since the interactions we have considered are not even the most general that
could be constructed with a Hilbert transform, there
may be many more weakly nonlocal Lorentz-violating interactions with
further interesting properties.

\section*{Acknowledgments}
The author is grateful to V. A. Kosteleck\'{y} for
helpful discussions.
This work is supported in part by funds provided by the U. S.
Department of Energy (D.O.E.) under cooperative research agreement
DE-FG02-91ER40661.

\end{document}